# Defect Regulation by Palladium Incorporation towards Grain Boundaries of Kesterite solar cells

Jinlin Wang[1,5]†, Jiangjian Shi[1]†, Kang Yin[1,5]†, Fanqi Meng[3]†, Shanshan Wang[4], Licheng Lou[1,5], Jiazheng Zhou[1,5], Xiao Xu[1,5], Huijue Wu[1], Yanhong Luo[1,5,6], Dongmei Li[1,5,6]*, Shiyou Chen[4]*, and Qingbo Meng[1,2,5,6]*

[1]Beijing National Laboratory for Condensed Matter Physics, Renewable Energy Laboratory Institute of Physics, Chinese Academy of Sciences (CAS); Beijing, 100190, P. R. China.

[2]Center of Materials Science and Optoelectronics Engineering, University of Chinese Academy of Sciences; Beijing, 100049, P. R. China.

[3]School of Materials Science and Engineering, Peking University; Beijing, 100871, P. R. China.

[4]Key Laboratory for Computational Physical Sciences (MOE), School of Microelectronics, Fudan University; Shanghai, 200433, P. R. China.

[5]School of Physical Sciences, University of Chinese Academy of Sciences; Beijing, 100049, P. R. China.

[6]Songshan Lake Materials Laboratory; Dongguan, 523808, P. R. China.

*Corresponding author. Email: dmli@iphy.ac.cn; chensy@fudan.edu.cn; qbmeng@iphy.ac.cn.

†These authors contributed equally to this work.

# Abstract

Kesterite $Cu_2ZnSn(S, Se)_4$ (CZTSSe) solar cell has emerged as one of the most promising candidates for thin-film photovoltaics. However, severe charge loss occurring at the grain boundaries (GBs) of Kesterite polycrystalline absorbers has hindered the improvement of cell performance. Herein, we report a redox reaction strategy involving palladium (Pd) to eliminate atomic vacancy defects such as $V_{Sn}$ and $V_{Se}$ in GBs of the Kesterite absorbers. We demonstrate that $PdSe_x$ compounds could form during the selenization process and distribute at the GBs and the absorber surfaces, thereby aiding in the suppression of Sn and Se volatilization loss and inhibiting the formation of $V_{Sn}$ and $V_{Se}$ defects. Furthermore, Pd(II)/Pd(IV) serves as a redox shuttle, i.e., on one hand, Pd(II) captures Se vapor from the reaction environment to produce $PdSe_2$, on the other hand, $PdSe_2$ provides Se atoms to the Kesterite absorber by being reduced to PdSe, thus contributing to the elimination of pre-existing $V_{Se}$ defects within GBs. These effects collectively reduce defects and enhance the p-type characteristics of the Kesterite absorber, leading to a significant reduction in charge recombination loss within the cell. As a result, high-performance Kesterite solar cells with a total-area efficiency of 14.5% have been achieved. This remarkable efficiency increase benefited from the redox reaction strategy offers a promising avenue for the precise regulation of defects in Kesterite solar cells and holds generally significant implications for the exploration of various other photovoltaic devices.

Kesterite $Cu_2ZnSn(S, Se)_4$ (CZTSSe) solar cell has emerged as one of the most promising candidates for thin-film photovoltaics, having garnered significant research attention in recent years ([1-4]). Encouragingly, there have been successive efficiency breakthroughs in this cell, with the most recent achievements reaching 13% and 14% ([5, 6]). However, the current efficiency still falls significantly short of the Shockley-Queisser limit. This efficiency gap, also observed as significant deficits in the open-circuit voltage ($V_{OC}$), are primarily caused by severe charge non-radiative recombination caused by defects in the CZTSSe absorber.

The CZTSSe absorber exhibits intricate defects owing to its composition of multinary elements and polycrystalline microstructures ([7-9]). Previous studies have predominantly focused on intrinsic point defects within the bulk grain interiors (GI), stemming from cation substitutions or disorders ([10-12]). However, recent research by Li et al. ([13]) highlighted that, in contrast to GI, grain boundaries (GBs) within CZTSSe absorbers actually play a more substantial role in influencing charge recombination velocity and charge loss. Several factors contribute to this phenomenon regarding GBs. Firstly, GBs typically exhibit a higher degree of structural distortion and atomic disorder, leading to the formation of various types of defects, such as Se-Se dimers ([14-16]). Secondly, conductive secondary phases such as $Cu_xSe$ and $SnSe_x$ tend to segregate at GBs ([17-19]), creating current shunting pathways within these boundaries. Thirdly, during the later stages of the selenization process, the low-Se-pressure reaction environment struggles to maintain CZTSSe crystals in a stable state, leading to gradual surface decomposition ([20, 21]). In particular, volatile Sn and Se elements can escape from both GBs and the film surfaces, resulting in the formation of Sn and Se vacancies (i.e., $V_{Sn}$ and $V_{Se}$) ([22, 23]), which are typically considered as detrimental non-radiative recombination-center defects.

Due to the aforementioned reasons, precise regulation of defects at GBs of CZTSSe absorbers has become an essential prerequisite for further enhancing the efficiency of Kesterite solar cells and consequently has garnered increasing attention. Researchers have developed a series of strategies for overcoming issues such as the detrimental Se-Se dimers and the segregation of secondary phases, by using alkali metal incorporation ([24, 25]), surface etching ([18, 26]), composition control ([27-29]), and optimization of reaction pathways ([6, 30]).

Furthermore, to suppress the vacancy defects at GBs, additional Sn sources (*27, 31-34*) have been introduced into the precursor or external selenization reaction environment. Oxygen/air post-treatment strategies have also been explored to passivate $V_{Se}$ defects by introducing $O_{Se}$ substitutions at GBs (*23, 35-37*). These efforts have yielded promising results sequentially; nonetheless, achieving comprehensive defect regulation in GBs remains a challenging endeavor in experimental settings.

Herein, we present a redox reaction strategy involving palladium (Pd) to mitigate the loss of volatile elements and associated vacancy defects in Ag-alloyed CZTSSe (ACZTSSe) absorber films. During the selenization process, Pd undergoes reactions to form $PdSe_x$ compounds (PdSe and $PdSe_2$), which are distributed in the GBs and surface regions, creating a heterogeneous coverage over the ACZTSSe grains within the film. These thermodynamically stable $PdSe_x$ compounds can effectively restrain the volatilization of Sn and Se elements from the GBs and the crystal surfaces of the film. Furthermore, the redox circles of Pd(II)/Pd(IV) provide a mechanism for filling the $V_{Se}$ in the GBs of the film. Pd(II) can readily capture vapor Se from the reaction environment, transforming into $PdSe_2$, which possesses a high oxidation capacity and subsequently can supply Se atoms to the ACZTSSe by being reduced to PdSe, thus eliminating $V_{Se}$ defects. With these advantages, our redox reaction strategy successfully stabilizes the chemical and electrical properties of the Kesterite film throughout the entire high-temperature selenization process. This results in low-defect-density ACZTSSe absorbers and high-performance Kesterite solar cells, achieving a total-area efficiency of 14.5% (certified at 14.3%).

**Influence of Pd on the selenization process**

In our experimental approach, we introduced $PdCl_2$ into the precursor solution for the subsequent fabrication of ACZTSSe films. The best performance was achieved when the Pd/Zn ratio was optimized at 1%. For clarity, we refer to the ACZTSSe film with Pd as ACZTSSe-Pd, while the Pd-free ACZTSSe is considered as the control sample. Through spherical aberration-corrected scanning transmission electron microscopy (STEM) and electron energy loss spectroscopy (EELS) characterization, we found that in the final-state ACZTSSe film, Pd element is primarily localized along the GB (Fig. 1A to B and fig. S1). A similar spatial distribution pattern was observed for the Se element in the GB regions (Fig. 1C and fig. S1). It suggests that the introduced Pd was mainly distributed in the GB regions, in the form of PdSe$_x$ compounds such as PdSe$_2$ and PdSe. Besides the GBs, Pd was also detected on the film surfaces by using X-ray photoelectron spectra (XPS), with higher intensity than that in the etched bulk region (fig. S2). This implies that PdSe$_x$ compounds mainly existed both at the GBs and on the film surfaces, forming effective heterogeneous coverings over the ACZTSSe grains. X-ray diffraction (XRD) and Raman investigations indicated that the incorporation of Pd did not alter the size and vibration properties of the ACZTSSe lattice, even when the Pd/Zn ratio was increased to 3% (fig. S3). Even during the high temperature selenization reaction process, Pd had neither altered the relative position of the XRD peaks of the ACZTSSe phase. Therefore, we propose that the doping of Pd atoms into the Kesterite lattice is negligible. In the ACZTSSe crystallization growth process, Pd was mainly segregated to the GBs and the surfaces, forming PdSe$_x$ coverings. This hypothesis is further supported by density functional theory (DFT) calculations, which shows a high formation energy of >1.0 eV for Pd/M (M=Cu, Zn, or Sn) substitutions in the Kesterite lattice (fig. S4).

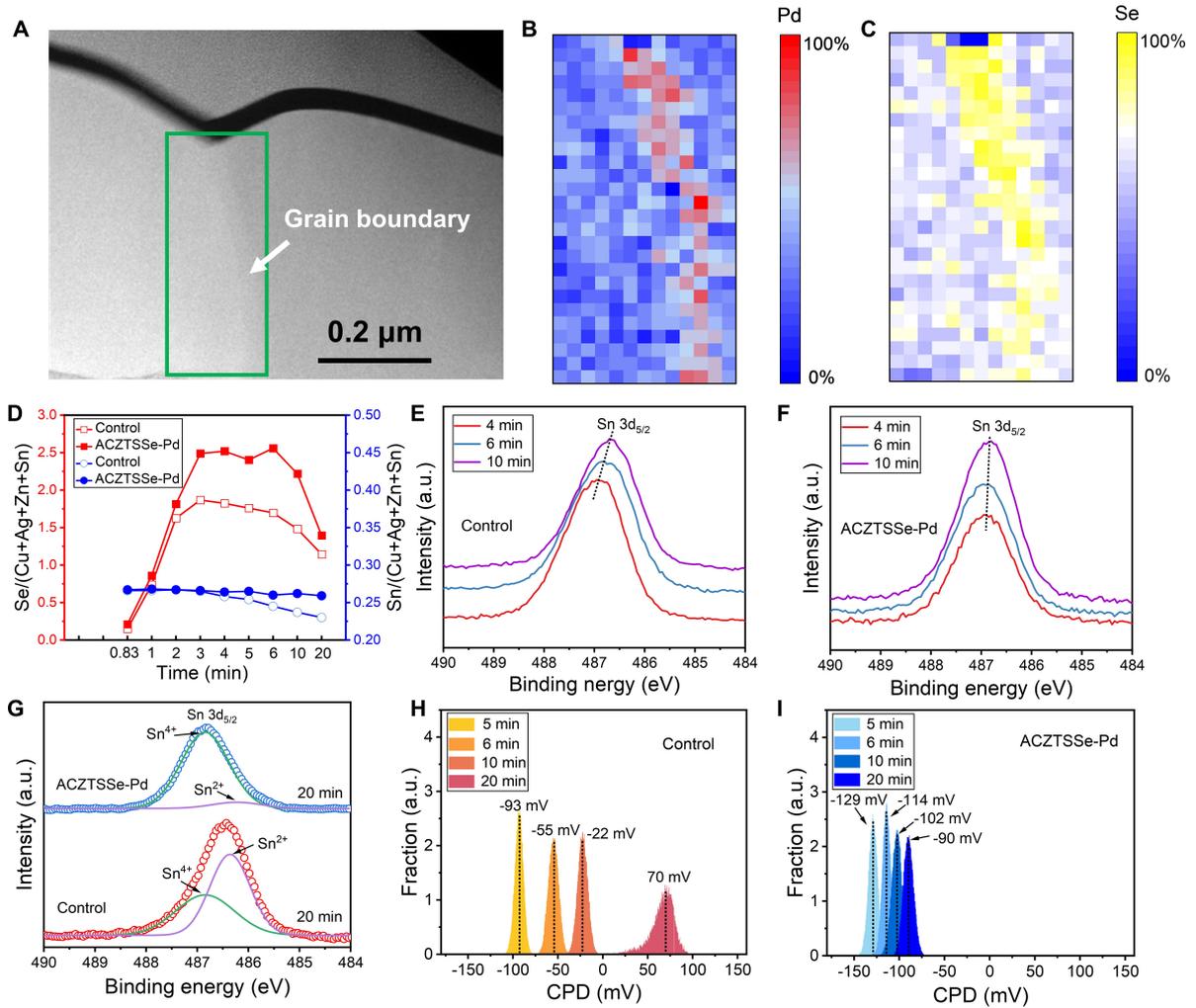

**Fig. 1. Influence of Pd on the selenization process.** (A) Cross-sectional HAADF STEM image of the ACZTSSe-Pd film. The green rectangle represents the EELS mapping region. (B-C) EELS mapping image of Pd (M4 and M5 peaks) and Se (L2 and L3 peaks). (D-F) Evolutions of element composition and Sn 3d XPS spectra of the films during the selenization process. (G) Gaussian fitting of the XPS spectra of the final-state ACZTSSe films. (H-I) Evolutions of surface CPD of the two samples during the selenization process.

We further delved into the impact of Pd on the selenization process of ACZTSSe films. We employed X-ray fluorescence (XRF) to quantify the element composition of films sampled at various selenization stages (Fig. 1D and fig. S5). It is evident that the volatilization loss of Sn and Se elements conspicuously and continuously occurred in the control sample when the selenization reaction exceeded 5 minutes. In contrast, the ACZTSSe-Pd sample effectively suppressed element loss, with the Sn/(Cu+Ag+Zn+Sn) ratio remaining nearly constant throughout the entire process. Additionally, a higher and more stable Se/(Cu+Ag+Zn+Sn) ratio was observed during the intermediate selenization process. This suppression of element loss consequently stabilized the electronic structure of Sn. The XPS peak position of Sn in the ACZTSSe-Pd sample remained almost unchanged as selenization proceeded (Fig 1E and 1G, and figs. S6 to S7). This is in stark contrast to the control sample, in which the Sn XPS peak continuously shifted towards the lower-energy direction (Fig 1F to G).

By further fitting, we found that the Sn XPS spectrum of the final-state control sample was composed of two peaks with comparable intensities, locating at 486.9 eV and 486.3 eV, which can be ascribed to Sn(IV) and Sn(II), respectively (Fig. 1G) (*38-40*). In contrast, the XPS spectrum of the ACZTSSe-Pd sample was dominated by the higher-energy peak (Fig. 1G). DFT calculations revealed that the appearance of Sn(II) is arisen from $V_{Se}$ defects, which results in lone pair electrons in the 5s orbit of the nearest neighboring Sn atom (figs. S8 to S9). As such, the continuous shift of Sn XPS spectra during the selenization process is indicative of the ongoing formation of $V_{Se}$, which is a donor defect that would weaken the p-type carriers of the Kesterite absorber (fig. S10) (*8, 41*). This result was confirmed by surface contacting potential difference (CPD) mapping of these films using Kelvin probe force microscopy (KPFM) (Fig. 1H and 1I and Fig. S11). For the control sample, its average CPD evolved from -93 mV to 70 mV as selenization progressed, resulting in an upshift of the Fermi energy level by more than 160 mV. In comparison, the CPD evolution of the ACZTSSe-Pd sample was limited to less than 40 mV (from -129 to -90 mV), demonstrating much more stable surface electrical properties and a more pronounced p-type nature.

**Characterization of ACZTSSe absorbers**

We further characterize the impact of Pd on the element distribution in final-state ACZTSSe absorber films using STEM based energy-dispersive X-ray spectroscopy (EDX) analysis. In the case of the control film (Fig. 2A to D and figs. S12 to S13), a clear deficiency of Sn and Se elements is evident in the GB region, while other elements display a relatively uniform distribution profile across the GBs. This observation indicates that Sn and Se elements in GB regions did indeed experience volatilization loss. In contrast, in the ACZTSSe-Pd film, the Se content in the GB region is slightly higher than that in the GI, and the distribution of Sn is uniform, similar to other elements (Fig. 2E to H and figs. S12 to S13). This suggests that the loss of Sn and Se elements in GB regions has been effectively mitigated. This result was further confirmed on a larger scale using scanning electron microscopy (SEM) (figs. S14 to S15).

CPD mapping results revealed that the suppression of element loss in GBs has significantly altered the energy band bending behaviors of the ACZTSSe absorber film. In the control film, obviously higher CPD was observed in the GB regions compared to the GIs (Fig. 2I), indicating a downward bending of the energy band (Fig. 2K). Under this case, the photo-generated minority electrons would be driven to the defective GB regions and cause carrier loss. The existence of high concentration of donor defects in the GBs is the primary reason for this phenomenon. In contrast, the ACZTSSe-Pd film exhibited an inversion of the energy band bending (Fig. 2J and 2L), benefiting from an effective reduction in the $V_{Se}$ donor defect. This upward bending would facilitate the spatial separation of minority electrons, thereby reducing carrier recombination.(*42, 43*). This was confirmed by transient photoluminescence (PL) measurements (fig. S16), which revealed that the ACZTSSe-Pd film exhibited a significantly prolonged carrier lifetime. Due to the reduced carrier recombination, the ACZTSSe-Pd film also displayed a higher steady-state PL intensity (fig. S17). Moreover, the ACZTSSe-Pd film exhibited a notably smaller PL bathochromic shift relative to its bandgap ($E_g$), more than 20 meV lower than that of the control sample (Fig.2M). This indicates a reduction in electrostatic potential fluctuations within the absorber film (*44-46*), which is also evidence of the suppressed defects.

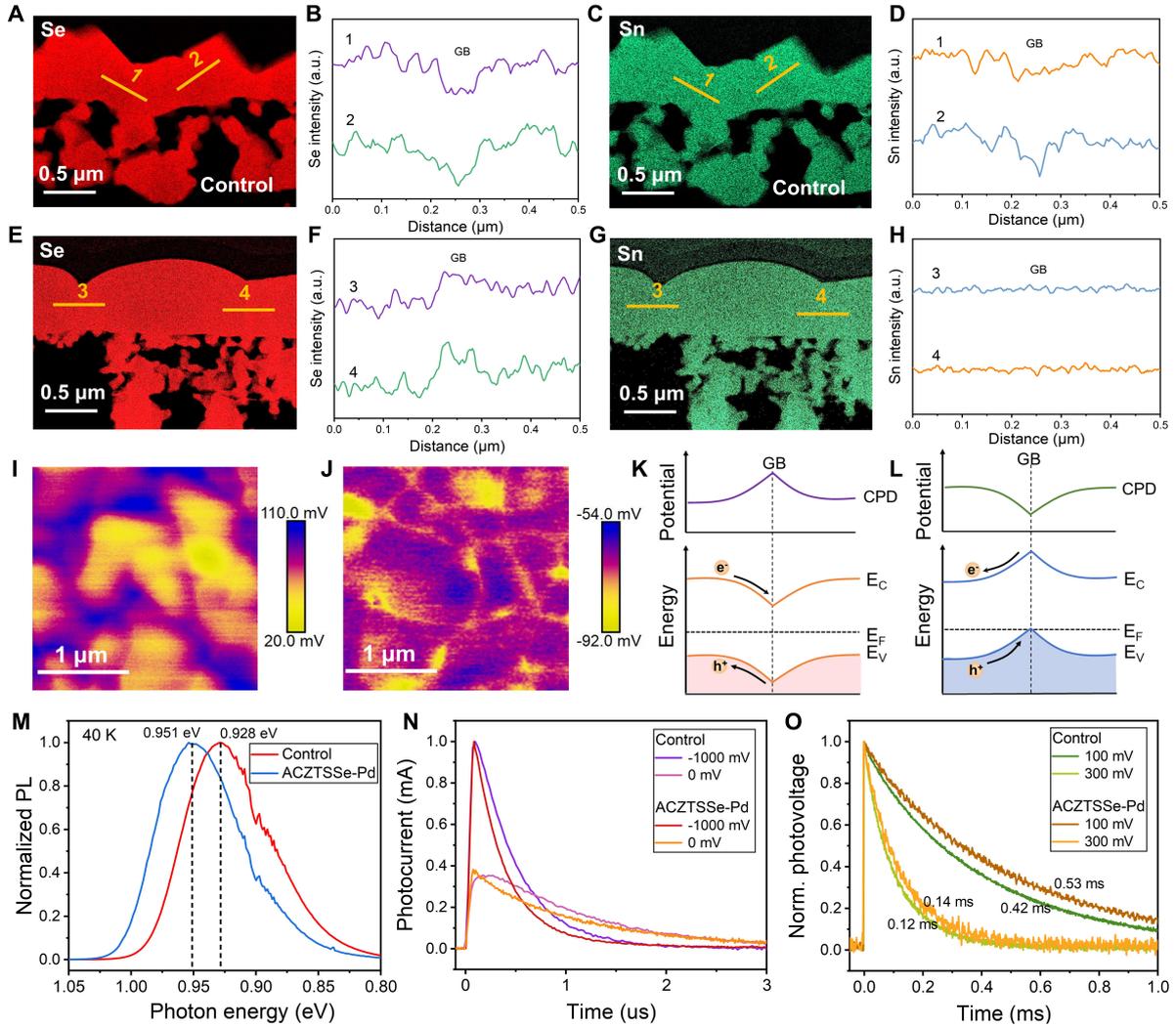

**Fig. 2. Characterization of the final-state ACZTSSe absorbers.** (A-H) Cross-sectional STEM-EDX mappings of Se and Sn elements in the films (A-B: Se, control film, C-D: Sn, control film, E-F: Se, ACZTSSe-Pd film, G-H: Sn, ACZTSSe-Pd). (I-L) KPFM mapping of the two absorbers and the schematic diagram of the energy band bending near the GB regions. "e$^-$" represents electron and "h$^+$" represents hole. (M) Steady-state PL spectra of the absorber films. (N-O) Modulated transient photocurrent and photovoltage dynamics of the cells.

These absorber films were subsequently used to fabricate solar cells following a standard device configuration (Fig. S18). We investigated the charge transport and recombination losses of the cells using modulated transient photocurrent/photovoltage measurements (M-TPC/TPV). It is evident that the ACZTSSe-Pd cell exhibited much faster photocurrent decays at both -1 and 0 V, with the photocurrent rise velocity and peak position remaining unaffected by the bias voltage (Fig. 2N). In contrast, these dynamics in the control cell were noticeably delayed when the built-in electric field within the cell was weakened by altering the bias voltage (Fig. 2N). These findings demonstrate that the impact of electrostatic potential fluctuations and carrier trapping processes on the charge transport have been effectively reduced in the ACZTSSe-Pd absorber (*47, 48*). These improvements also suppressed charge recombination, as evidenced by the slower photovoltage decay observed in the cell under different positive bias voltages (Fig. 2O).

**Device performance and characterization**

The top-performing cell achieved an impressive power conversion efficiency (PCE) of 14.5%, with a short-circuit current density ($J_{SC}$) of 36.7 mA cm$^{-2}$, a $V_{OC}$ of 0.555 V, and a fill factor (FF) of 0.712 (Fig. 3A). In comparison, the control cell exhibited a lower PCE of only 12.8%, with noticeably lower $J_{SC}$ at 35.5 mA cm$^{-2}$, $V_{OC}$ at 0.519 V, and FF at 0.695. A statistical analysis of these parameters further highlighted the performance disparity between these two cell types (fig. S19). External quantum efficiency (EQE) spectra indicated that the absorbers in both cells had a similar bandgap of 1.1 eV. The improvement in $J_{SC}$, approximately 1 mA cm$^{-2}$, primarily originated from EQE enhancement in the wavelength range from 600 to 1080 nm (fig. S20). This suggests that photocarriers generated within the bulk absorber were more effectively extracted (*4, 49*), due to reduced charge losses within the GBs. The reduction in charge losses is also responsible for the significant improvements in $V_{OC}$ and FF.

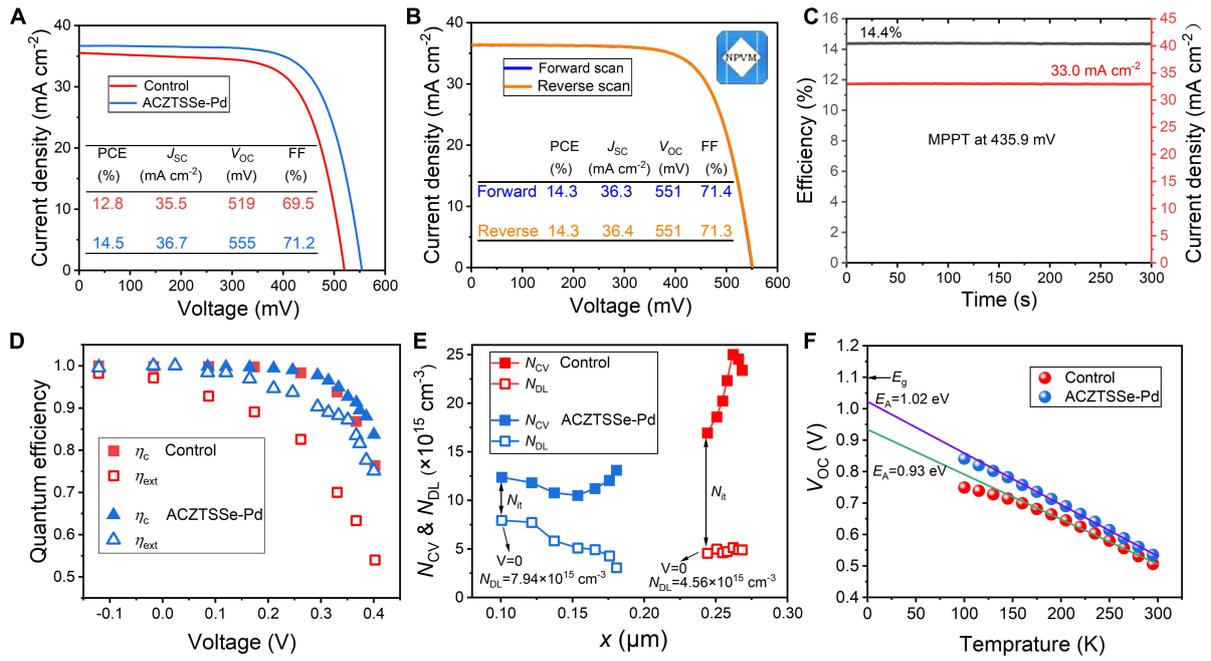

**Fig. 3. Device characterization.** (A) Current-voltage (*I-V*) characteristics of the champion control and ACZTSSe-Pd devices. (B) Certified I-V curves of the cell at both forward and reverse scanning directions. (C) Maximum power point tracking of the cell for 300 s. (D) Charge extraction and collection efficiencies of the cells derived from modulated electrical transient measurements. (E) Charge profiles of the cells measured by DLCP and C-V. (F) Temperature-dependent $V_{OC}$ of the cells.

Moreover, the cell has received a certified PCE of 14.3% from an accredited independent testing laboratory, the National PV Industry Measurement and Testing Center (NPVM) (Fig. 3B and fig. S21). The maximum power point tracking (MPPT) of the cell was also conducted in the certification process. When biased at 435.9 mV, the cell gave constant current output of about 32.95 mA cm$^{-2}$ for 300 s, achieving a steady-state PCE of approaching 14.4% (Fig. 3C). This achievement stands as the highest result reported to date.

We further quantified the charge loss in the cell based on the modulated electrical transient measurements (Fig. 3D). The primary difference between these two cells lies in the charge extraction efficiency ($\eta_{ext}$), which is correlated to the charge loss in the bulk absorber. At 0.4 V, the ACZTSSe-Pd solar cell achieved a 1.4-fold enhancement in $\eta_{ext}$. Furthermore, the ACZTSSe-Pd solar cell also exhibited higher charge collection efficiency ($\eta_C$) at voltages exceeding 0.35 V. This indicates that interface defect-induced charge loss in the cell has also been reduced. The findings from the electrical transient analysis are further substantiated by a direct measurement of charge spatial distribution within the cell employing capacitance-voltage (*C-V*) and drive-level capacitance profiling (DLCP) methods (Fig. 3E). Based on the difference in charge density measured by DLCP and *C-V (50)*, the interface defect density ($N_{IT}$) of the ACZTSSe-Pd solar cell is estimated to be $4.4×10^{15}$ cm$^{-3}$, which is only approximately one-third of that observed in the control solar cell ($1.2×10^{16}$ cm$^{-3}$). Furthermore, the ACZTSSe-Pd absorber exhibited a higher charge density at 0 V, indicative of improved p-type doping. This aligns well with the CPD mapping results and serves as direct evidence of the suppressed $V_{Se}$ donor defect in the surface region. The enhancement of film surface quality is also evident from temperature-dependent $V_{OC}$ measurements (*2, 51*). In the ACZTSSe-Pd device, the activation energy ($E_A$) was determined to be 1.02 eV, closely matching the optical $E_g$ of the absorber (1.1 eV), whereas the control device exhibited a considerably lower $E_A$ of 0.93 eV (Fig. 3F).

**Microscopic mechanism of the Pd assisted GB engineering**

We proceeded to explore the microscopic mechanisms underlying the defect regulation at GBs after Pd incorporation. To ascertain the form in which PdSe$_x$ compounds existed, we measured XPS spectra of the Pd element throughout the selenization process. In the precursor film, Pd predominantly exhibited a +2 valence state (fig. S22). During the intermediate selenization reaction stages (at 4 minutes and 6 minutes), a proportion of Pd appeared in the +4 valence state (Fig. 4A), indicating the formation of PdSe$_2$, while other Pd ions remained in the +2 valence state, possibly as PdSe. These PdSe$_x$ compounds were also corroborated through XRD and Raman characterization (figs. S23 to S24). Through DFT calculations, we discovered that PdSe$_2$ and PdSe possess higher cohesive energy than the corresponding SnSe$_x$ compounds (Fig. 4B), signifying superior compound stability and lower volatility. Consequently, from a reaction dynamics perspective, the PdSe$_x$ heterogeneous coverage could mitigate element volatilization from the GBs and surfaces of the ACZTSSe absorber by creating a locally saturated environment. This, in turn, reduces the formation of $V_{Sn}$ and $V_{Se}$ defects.

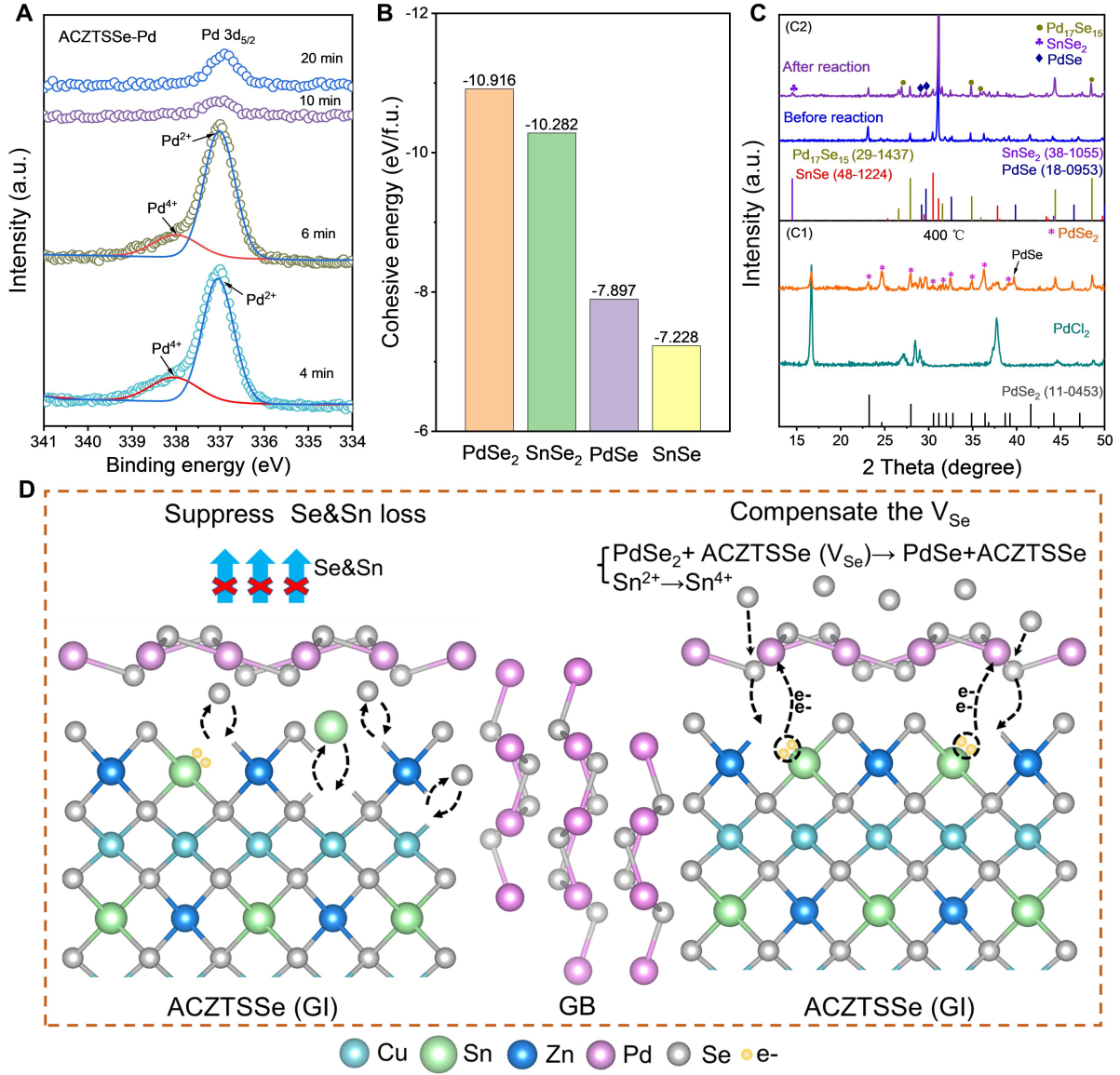

**Fig. 4. Microscopic mechanism of the Pd assisted defect regulation in the GB regions.** (A) XPS spectra of the Pd element during the selenization process. (B) Cohesive energies of $PdSe_2$, $SnSe_2$, PdSe and SnSe. (C) XRD patterns of $PdCl_2$, its selenization product $PdSe_2$ (C1), and the reaction products of $PdSe_2$ and SnSe (C2). (D) Schematic diagram of possible roles that Pd played in the selenization process. Left: preventing element volatilization loss through heterogeneous coverings; right: compensating the $V_{Se}$ defect through Se atom exchange between reaction environment, $PdSe_2$ and ACZTSSe film with Se vacancies.

Moreover, during the later stages of selenization, XPS results indicated a disappearance of the +4 valent Pd, with Pd predominantly remaining in the +2 valence state (Fig. 4A). This phenomenon was further validated by XRD and Raman spectra (figs. S23 to S24). The valence state evolution of Pd suggests the occurrence of redox reactions. Firstly, through selenization conducted at different temperatures (as described in Supplementary Text 1), we observed that the oxidation of Pd(II) to Pd(IV), leading to the formation of $PdSe_2$, readily occurred even at relatively low temperatures and under low Se vapor partial pressure (Fig. 4C1 and fig. S25). Secondly, we discovered that $PdSe_2$ exhibits a high propensity for oxidation when it is mixed with SnSe at elevated temperatures (as detailed in Supplementary Text 2). XRD analysis indicated that SnSe would undergo oxidation to form $SnSe_2$, while $PdSe_2$ was reduced to form $PdSe_x$ compounds, such as PdSe and $Pd_{17}Se_{15}$ (Fig. 4C2), as described by the reaction:

$$PdSe_2 + SnSe \rightarrow PdSe_x + SnSe_2. \tag{1}$$

The occurrence of this redox reaction was substantiated through DFT calculations, which revealed an enthalpy change of -3.3 kJ/mol at $x=1$. Furthermore, the notable disparity in standard electrode potential ($E^\theta$) between Pd(IV)/Pd(II) ($E^\theta > 1$ V) and Sn(IV)/Sn(II) ($E^\theta < 0.2$ V) provided additional support for this redox reaction (*52*). These findings imply that $PdSe_2$ can assist in maintaining Sn in a +4-valence state by supplying Se atoms and capturing excess electrons. In our view, this redox mechanism is also applicable to the mixing system comprising $PdSe_2$ and the ACZTSSe containing low-valence Sn and $V_{Se}$, as follows:

$$PdSe_2 + ACZTSSe\ (V_{Se}) \rightarrow PdSe + ACZTSSe. \tag{2}$$

This scenario is reasonable because the electron and Se transfer in reaction (2) closely resembles that in the confirmed reaction (1), primarily occurring between Pd(IV) and Sn(II). It is evident that the occurrence of reaction (2) will assist in mitigating the $V_{Se}$ defects that have already formed in the GB regions and on the surfaces of ACZTSSe absorbers. For clarity, we have schematically depicted the microscopic mechanism of Pd-assisted defect suppression in the Kesterite absorber in Fig. 4D. First, the $PdSe_x$ compounds formed during the selenization process act as a heterogeneous covering layer of the GBs and the absorber's surface. This effectively suppresses the Sn and Se element volatilization, preventing the formation of $V_{Sn}$ and $V_{Se}$ defects. Second, Pd(II)/Pd(IV) functions as a redox shuttle, capturing vapor Se from the reaction environment to form $PdSe_2$. Subsequently, $PdSe_2$ can provide Se atoms to the ACZTSSe absorber by being reduced to PdSe, thereby aiding in the elimination of $V_{Se}$ defects. Overall, this redox reaction mechanism offers a promising avenue for the precise regulation of defects in Kesterite solar cells, and also holds implications for the GB engineering in other photoelectric devices.

**Acknowledgments:** The authors acknowledge the Natural Science Foundation of China (grants U2002216, 52222212, 51972332, 52172261, 52227803).

**Author contributions:** Jinlin Wang, Jiangjian Shi, Dongmei Li and Qingbo Meng conceived the idea and designed the experiments. Jinlin Wang and Jiangjian Shi did the experiments and the data analysis. Licheng Lou, Jiazheng Zhou and Xiao Xu supported CZTSSe solar cell fabrication. Fanqi Meng and Kang Yin performed STEM and DLCP measurements. Huijue Wu and Yanhong Luo supported M-TPC/TPV characterization and discussions. Jinlin Wang, Jiangjian Shi, Dongmei Li and Qingbo Meng participated in writing the manuscript. Shanshan Wang and Shiyou Chen provided DFT results and valuable discussion to the work.

**Competing interests:** The authors declare no competing financial or nonfinancial interests.